\documentclass[aps,epsf,epsfig,twocolumn]{revtex4}
\usepackage[dvips]{graphicx}
\usepackage[usenames]{color}
\def\be{\begin{equation}}
\def\ee{\end{equation}}
\def\bea{\begin{eqnarray}}          
\def\eea{\end{eqnarray}}
\def\bi{\begin{itemize}}
\def\ei{\end{itemize}}
\def\bin{\begin{enumerate}}
\def\ein{\end{enumerate}}
\def\la{\langle}
\def\ra{\rangle}

\begin{document}

\title{$N$-conserving Bogoliubov vacuum of a two component Bose-Einstein
condensate: Density fluctuations close to a phase separation condition}

\author{ Bart\l{}omiej Ole\'s and Krzysztof Sacha }

\address{Marian Smoluchowski Institute of Physics and 
Mark Kac Complex Systems Research Center,
          Jagiellonian University,
          Reymonta 4, 30-059 Krak\'ow, Poland  }

\date{\today}

\begin{abstract}
Two component Bose-Einstein condensates are considered within a number
conserving version of the Bogoliubov theory. We show that the Bogoliubov vacuum
state can be obtained in the particle representation in a simple form. 
We predict considerable density fluctuations in finite systems close to the
phase separation regime. We analyze homogeneous condensates and condensates in a double
well potential.
\end{abstract}

\maketitle

\section{Introduction}

Bose-Einstein condensate (BEC) is a unique state of a many particle system where, ideally,
all particles occupy the same single particle state. It is obviously possible for bosons
only, and experimentally it can be realized in ultra-cold dilute atomic gases
\cite{review1}. Since the
first experimental realization numerous different phenomena involving BEC have been
investigated 
and nowadays it is also possible to obtain mixtures of BECs or even mixtures of
ultra-cold bosonic and fermionic gasses \cite{twocomp}. 
Two-component BEC \cite{twocomp} can reveal 
number of interesting phenomena, e.g., phase separation
\cite{phasesep,timm98}, self-localization \cite{selfloc}, 
condensate entanglement \cite{entnature,sorensen} or internal Josephson effects
\cite{timm03}.

In an infinite homogeneous system the phase separation occurs abruptly
once interactions reach their critical values \cite{timm98,timm03}. In the present paper
we show effects of a finite system. That is, in a finite box there is a region 
close to critical values of the coupling parameters where substantial density 
fluctuations can be observed.

A standard theoretical description of a single condensate and condensate mixtures starts with 
the mean field Gross Pitaevskii equations \cite{GPEcite} that 
provide estimates for ground states and collective 
excitations of a system but under an assumption that it is described by
perfect condensate product states. 
Particle interactions, however, can lead
to substantial depletion of the condensates \cite{review1} and
in order to obtain a more realistic picture usually a Bogoliubov theory is applied,
which allows one to describe small quantum corrections to the mean field solution
\cite{review1,castin,BTcite,gardiner}. The key idea of the original Bogoliubov 
theory \cite{BTcite} (usually used in the BEC field) 
is the $U(1)$ symmetry breaking approach where 
the atomic field operator is assumed to have a nonzero expectation value. 
This {\it coherent} state necessarily involves superposition of states 
with different numbers of atoms, an assumption very far from experimental
reality. Moreover, careful analysis of the original theory shows that
the Bogoliubov-de~Gennes equations correspond to an eigenvalue problem
of an operator which is not diagonalizable and the theory must break down
after a finite time \cite{castin,lewyou}.

To overcome these drawbacks we 
employ a number conserving version of the Bogoliubov theory, which has been presented 
by Castin and Dum \cite{castin} (see also \cite{gardiner}) for a one-component BEC
and generalized to a two-component system by S{\o}rensen \cite{sorensen}, 
and analyze the Bogoliubov vacuum state of a mixture of two BECs.
The two theories should give the same physical predictions for large particle numbers.
There are, however, examples of systems where the $N$-conserving theory works in a regime of 
the standard theory breakdown \cite{dzsacha03}.
The Bogoliubov vacuum is usually obtained in the quasi-particle representation where
quantum depletion, i.e. the number of particles occupying non-condensate modes can be easily
calculated for a given system \cite{review1,castin,BTcite,gardiner}.
To gain insight into the form of the ground state of the system,
we derive the Bogoliubov ground state in the particle representation.
This enables us to
perform simulations
of density measurements in single experiments \cite{Fock,dziarmaga,trippen}.

For a single condensate the Leggett ansatz of the vacuum for translationally invariant systems 
\cite{ansatzL} has been shown to be valid in any inhomogeneous condensates in 
\cite{dziarmaga,dzsacha03}. In the present paper we show that the ansatz can be used also
in the two-component case even in the presence of the inter-species interaction. 
The obtained Bogoliubov vacuum state is then used in an analysis of density fluctuations
in 3D homogeneous condensates and in condensates trapped in a double well potential. 
It turns out that vicinity of the critical point for the phase separation is especially 
interesting because the fluctuations there become considerable.

The paper is organized as follows. In Sec.~II
we present the solution for the Bogoliubov vacuum state in the
particle representation, derived within the
number conserving version of the Bogoliubov theory.
In Sec.~III we describe a procedure
used later to perform density measurement simulations.
The theory is applied to the analysis of
homogeneous condensates in Sec.~IV and to the double well problem in Sec.~V.
We conclude in Sec.VI. 
Short reminder of the Bogoliubov theory \cite{sorensen} is presented in the Appendix A
and details of the derivation of the Bogoliubov vacuum state in the particle representation 
are presented in the Appendix B.

\section{Bogoliubov vacuum state} 
\label{BVS} 

We consider a two component Bose-Einstein condensate formed by a mixture 
of two kinds of atoms (or the same atoms in two different internal
states), i.e. $N_a$ atoms of type $a$ and $N_b$ atoms
of type $b$. The Hamiltonian of the system reads
\bea
\hat H&=& \int d^3r\left( \hat\psi_a^\dagger\left[-\frac{\hbar^2}{2m_a}
\nabla^2+V_a(\vec r)+\frac{g_{a}}{2}\hat\psi_a^\dagger\hat\psi_a\right]
\hat\psi_a\right.\cr
&&+\hat\psi_b^\dagger\left[-\frac{\hbar^2}{2m_b}
\nabla^2+V_b(\vec
r)+\frac{g_{b}}{2}\hat\psi_b^\dagger\hat\psi_b\right]\hat\psi_b\cr
&&\left.+g\hat\psi_a^\dagger\hat\psi_b^\dagger\hat\psi_a\hat\psi_b \right),
\eea
where $m_a$, $m_b$ are particle masses, $V_a(\vec r)$, $V_b(\vec r)$ stand for 
the trapping potentials and
\bea
g_{a}&=&\frac{4\pi\hbar^2a_a}{m_a}, \cr
g_{b}&=&\frac{4\pi\hbar^2a_b}{m_b}, \cr
g&=&2\pi\hbar^2a_{ab}\left(\frac{1}{m_a}+\frac{1}{m_b}\right),
\eea
where $a_a$, $a_b$, $a_{ab}$ are the scattering lengths.
The number conserving Bogoliubov theory \cite{sorensen,castin} assumes the following decomposition
of the bosonic field operators
\bea
\hat\psi_a(\vec r)&=&\phi_{a0}(\vec r)\hat a_0+\delta\hat\psi_a(\vec r), \cr
\hat\psi_b(\vec r)&=&\phi_{b0}(\vec r)\hat b_0+\delta\hat\psi_b(\vec r),
\label{decompR}
\eea
where we separate the operators $\hat a_0$ and $\hat b_0$ that annihilate
atoms in modes $\phi_{a0}$ and $\phi_{b0}$, respectively,
which are macroscopically occupied by atoms. That is, for the
states we are after 
\bea
\la \hat a_0^\dagger\hat a_0\ra \approx N_a, &&
\la \hat b_0^\dagger\hat b_0\ra \approx N_b. 
\eea
Corrections $\delta\hat\psi_a$ and
$\delta\hat\psi_b$ are thus supposed to be small and we may perform expansion
of the Hamiltonian in powers of $\delta\hat\psi_a$ and $\delta\hat\psi_b$.
In the zero order, condition for energy extremum in the $\phi_{a0}$ and 
$\phi_{b0}$ space leads to coupled Gross-Pitaevskii equations
\bea
H_{GP}^a\phi_{a0}=0, && H_{GP}^b\phi_{b0}=0,
\label{GPER}
\eea
where
\bea
H_{GP}^a&=&-\frac{\hbar^2}{2m_a}
\nabla^2+V_a+g_{a}N_a|\phi_{a0}|^2+gN_b|\phi_{b0}|^2-\mu_a, \cr
H_{GP}^b&=&-\frac{\hbar^2}{2m_b}
\nabla^2+V_b+g_{b}N_b|\phi_{b0}|^2+gN_b|\phi_{b0}|^2-\mu_b, \cr
&&
\eea
(with chemical potentials $\mu_a$ and $\mu_b$)
that allow us to find single particle modes macroscopically 
occupied by atoms.
The first order terms of the Hamiltonian disappear. In the second order one
obtains an effective Hamiltonian which, employing the Bogoliubov
transformation, can be written in a diagonal form
\be
\hat H_{\rm eff}\approx\sum_{n\in"+"} E_n \hat c_{n}^\dagger\hat c_n,
\label{efffin}
\ee
where the sum goes over the so-called
family "+" solution of the Bogoliubov equations (see Appendix A).
The quasi-particle annihilation operators are defined as:
\bea
\hat c_n&=& \langle u_n^a | \hat\Lambda_a \rangle - \langle v_n^a | \hat\Lambda_a^\dagger \rangle
+ \langle u_n^b | \hat\Lambda_b \rangle - \langle v_n^b | \hat\Lambda_b^\dagger
\rangle, 
\label{boper}
\eea
where
\bea
\hat\Lambda_a(\vec r)&=&\frac{\hat a_0^\dagger}{\sqrt{N_a}}\delta\hat\psi_a(\vec
r), \cr
\hat\Lambda_b(\vec r)&=&\frac{\hat b_0^\dagger}{\sqrt{N_b}}\delta\hat\psi_b(\vec
r).
\label{Lambdaop}
\eea
The wavefunctions $\{u_n^a, v_n^a, u_n^b, v_n^b \}$ 
are solutions of the Bogoliubov equations corresponding to eigenvalue $E_n$
(see Appendix A). Let us now switch to our results.

The Bogoliubov vacuum state $|0_B\ra$ is an eigenstate of the effective
Hamiltonian that is annihilated by all quasi-particle annihilation operators,
\be
\hat c_n|0_B\ra=0.
\ee
Other eigenstates can be generated by acting with the quasi-particle
creation operators $\hat c_n^\dagger$ on the Bogoliubov vacuum. 
The quasi-particle representation is thus natural to represent the system
eigenstates within the Bogoliubov theory. It is also suitable to obtain 
low order correlation functions.
However, to get predictions for density measurements, i.e. to simulate
measurements of all atom positions, the particle representation
turns out to be much more convenient.

In the Appendix B we show that the Bogoliubov vacuum state 
can be written in the particle representation in the following simple from 
\bea
|0_B\ra &\sim& 
\left[\left(\hat a_0^\dagger\right)^2+\sum_{\alpha=1}^\infty 
\lambda_{\alpha}^a\left(\hat a_\alpha^\dagger\right)^2\right]^{N_a/2} \cr
&&\times 
\left[\left(\hat b_0^\dagger\right)^2+\sum_{\alpha=1}^\infty 
\lambda_{\alpha}^b\left(\hat b_\alpha^\dagger\right)^2\right]^{N_b/2} |0\ra
\label{final}
\eea
where 
\bea
\lambda_{\alpha}^a=
\frac{\la \phi_{a\alpha}|\hat\Gamma_a|\phi_{a\alpha}^*\ra}{dN^a_\alpha+1}, 
\quad
\lambda_{\alpha}^b=
\frac{\la \phi_{b\alpha}|\hat\Gamma_b|\phi_{b\alpha}^*\ra}{dN^b_\alpha+1}.
\eea
The particle creation operators $\hat a_\alpha^\dagger$ ($\hat
b_\alpha^\dagger$) create particles in modes $\phi_{a\alpha}$ ($\phi_{b\alpha}$)
that are eigenstates of the single particle density matrices, and 
$dN^a_\alpha$ ($dN^b_\alpha$) are the corresponding eigenvalues, i.e.
\bea
\la 0_B| \hat\psi_a^\dagger(\vec r) \hat\psi_a(\vec r\;')|0_B\ra
&\approx& N_a\phi_{a0}^*(\vec r)\phi_{a0}(\vec r\;') \cr
&+&\sum_{\alpha=1}^\infty dN^a_\alpha
\phi_{a\alpha}^*(\vec r)\phi_{a\alpha}(\vec r\;'), 
\eea
and similarly for 
$\la 0_B |\hat\psi_b^\dagger(\vec r) \hat\psi_b(\vec r\;')|0_B\ra$.
The operators $\hat\Gamma_a$ and $\hat\Gamma_b$ are defined as:
\bea
\hat\Gamma_a=\sum_{n\in"+"}|u_n^a\ra\la v_n^a|, \quad
\hat\Gamma_b=\sum_{n\in"+"}|u_n^b\ra\la v_n^b|.
\label{gammaoper}
\eea
The presented solution (\ref{final}) is 
self-consistent provided the $\hat\Gamma_{a,b}$ 
operators are diagonal in the basis of the eigenvectors of the single 
particle density matrices.
In the following sections we show examples of a spatially
homogeneous system and BECs in a double well potential, 
where this indeed is the case.

\section{Density measurement}

Average particle density corresponds to the reduced single particle density which can be 
easily calculated within the Bogoliubov theory \cite{dziarmaga}.
The average density means an averaged
picture obtained by collecting outcomes of the density measurement in many 
experimental realizations of a system in the same quantum state. Even at zero temperature
a many body system, 
can reveal density fluctuations 
and a single photo of the system may be significantly different from the averaged picture 
\cite{Fock,dziarmaga,trippen}.

In order to perform density measurement simulations
we generally need a full many body probability density. As the number of particles
grows, however, using this density quickly becomes a very formidable task.
Instead one may use a sequential method proposed by Javanainen and Yoo \cite{Fock}.  
It relies on a choice of a position 
of a subsequent atom with the help of a conditional density probability which takes it into
account that previous atoms have already been found at certain positions. 
Note that, since this method requires acting with particle annihilation 
operators on the Bogoliubov vacuum, using the Bogoliubov state in
the quasi-particle representation would require inversion of the nonlinear
transformation (\ref{boper}). Having the state (\ref{final}) we avoid this problem.

In practice the sequential method \cite{Fock} can be used if only one (or few)
non-condensate modes are important. If many modes are relevant 
we should e.g. switch to an approximate method \cite{dziarmaga}. 
Suppose there are $M_a$ and $M_b$ 
modes where 
\be
\Lambda^{a,b}_\alpha\equiv\frac{|\lambda^{a,b}_\alpha|}{1-|\lambda^{a,b}_\alpha|} \gg 1.
\ee 
Then results of the density measurements 
corresponding to a state of the form (\ref{final}) can be approximated by 
\cite{dziarmaga}
\bea
\sigma_a(\vec r)&\sim&\left|\phi_{a0}(\vec
r)+\frac{1}{\sqrt{N_a}}\sum_{\alpha=1}^{M_a}
q_{a\alpha}\;\varphi_{a\alpha}(\vec r)\right|^2, \cr
\sigma_b(\vec r)&\sim&\left|\phi_{b0}(\vec
r)+\frac{1}{\sqrt{N_b}}\sum_{\alpha=1}^{M_b}
q_{b\alpha}\;\varphi_{b\alpha}(\vec r)\right|^2,
\label{sigmy}
\eea
where 
\bea
\varphi_{a\alpha}(\vec r) &=& \phi_{a\alpha}(\vec r)\; e^{-i{\rm Arg}(\lambda^a_\alpha)/2},
\cr
\varphi_{b\alpha}(\vec r) &=& \phi_{b\alpha}(\vec r)\; e^{-i{\rm Arg}(\lambda^b_\alpha)/2},
\label{phases}
\eea
and real parameters 
$q_{a\alpha}$ and $q_{b\alpha}$ have to be chosen randomly, for each 
experimental realization, according to a Gaussian probability density
\be
P(q_a,q_b)\sim \prod_{\alpha=1}^{M_a} 
\exp\left(-\frac{q_{a\alpha}^2}{\Lambda^a_\alpha}\right) 
\prod_{\beta=1}^{M_b}\exp\left(-\frac{q_{b\beta}^2}{\Lambda^b_\beta}\right).
\label{probd}
\ee
The replacement (\ref{phases}) is essential because it makes all eigenvalues 
of the $\hat\Gamma_{a,b}$ operators non-negative and allows writing the Bogoliubov
vacuum state of the form (\ref{final}) as a Gaussian superposition 
over condensates which, in turn, leads directly to the predictions 
(\ref{sigmy}) \cite{dziarmaga}.


\section{Homogeneous condensates}
\label{homopart}

The two component homogeneous condensate is an example of a Bose system 
where the Bogoliubov theory gives analytical results even in the
presence of a process which transfers atoms between the two components
(Rabi or Josephson coupling \cite{timm03}).
In numerous papers the
quasi-particle excitation spectrum is analyzed as well as its dynamical 
instability leading to the phase separation \cite{homog,timm98,timm03}.
In the present publication we fix the number of atoms in each component
and study the Bogoliubov vacuum state in the particle representation
for interaction parameters approaching the phase separation condition.

Suppose we deal with a condensate mixture in a box of $L\times L\times L$ 
size with periodic boundary conditions and all interactions are of
repulsive character, i.e. $g_a$, $g_b$, $g>0$.
The ground state solution of the 
Gross-Pitaevskii equations reveals condensate wavefunctions
\be
\phi_{a0}=\frac{1}{\sqrt{L^3}}, \quad
\phi_{b0}=\frac{1}{\sqrt{L^3}},
\label{hgpesol}
\ee
and chemical potentials
\be
\mu_{a}=g_a \rho_a+g \rho_b, \quad
\mu_{b}=g_b \rho_b+g \rho_a,
\ee
where $\rho_{a,b} = N_{a,b}/L^3$ are densities of $a$ and $b$ components.
For a homogeneous system it is appropriate to switch to the momentum space and
look for the solution of the Bogoliubov-de~Gennes equation in the form
\be
\left(
\begin{array}{c}
u_{k}^a  \\
v_{k}^a \\
u_{k}^b \\
v_{k}^b
\end{array}
\right)\frac{e^{i\vec k\cdot\vec r}}{\sqrt{L^3}}.
\ee
Then, one obtains two quasi-particles for each $\vec k$ 
with energies \cite{homog,timm98,timm03},
\bea
	E_{k,\pm}&=& \left[ \frac{\omega^2_{ak} + \omega^2_{bk}}{2} \right. \cr
		 &&\pm \left. \sqrt{ 
		 \frac{\left(\omega^2_{ak} -\omega^2_{bk}\right)^2}{4}
		+ \frac{\hbar^2 k^4}{m_a m_b} g^2 \rho_a \rho_b}
		\right]^{1/2},
\label{hbogspec}
\eea
where 
\bea
	\omega^2_{ak}&=& \frac{\hbar^2 k^2}{2 m_a} \left( \frac{\hbar^2 k^2}{2 m_a} 
		+ 2 g_a \rho_a \right) , \cr
	\omega^2_{bk}&=& \frac{\hbar^2 k^2}{2 m_b} \left( \frac{\hbar^2 k^2}{2 m_b}
		+ 2 g_b \rho_b \right),
\eea
and modes
\be
\left(
\begin{array}{c}
	u_{k,\pm}^a \\ 
	v_{k,\pm}^a \\
	u_{k,\pm}^b \\
	v_{k,\pm}^b
\end{array}
\right)= 
\left(
\begin{array}{c}
	2g {\cal E}_{kb} \left({\cal E}_{ka}+E_{k,\pm}\right) \sqrt{\rho_a \rho_b}\\ 
	2g {\cal E}_{kb} \left({\cal E}_{ka}-E_{k,\pm}\right) \sqrt{\rho_a \rho_b}\\ 
	\left(E^2_{k,\pm} - \omega^2_{ak}\right)\left({\cal E}_{kb}+E_{k,\pm}\right)\\
	\left(E^2_{k,\pm} - \omega^2_{ak}\right)\left({\cal E}_{kb}-E_{k,\pm}\right)
\end{array}
	\right) \chi_\pm,
\ee
where
\be
{\cal E}_{ka}=\frac{\hbar^2 k^2}{2 m_{a}}, \quad
{\cal E}_{kb}=\frac{\hbar^2 k^2}{2 m_{b}},
\ee
and the normalization factor
\bea
	\chi_\pm = \left\{ {4 {\cal E}_{kb} \left[4 {\cal E}_{ka} 
		{\cal E}_{kb} g^2 \rho_a \rho_b 
		+ \left(E_{k,\pm}^2 - \omega^2_{ak}\right)^2\right]E_{k,\pm}}
		\right\}^{-1/2}. \cr
\eea
Note that in the finite box the momenta are discrete
\be
\vec k=\frac{2\pi}{L}\left(n_x\vec e_x + n_y\vec e_y + n_z\vec e_z\right),
\ee
where $n_x,n_y,n_z$ are non-zero integers.

The reduced single particle density 
matrices are diagonal in the $e^{i\vec k\cdot \vec
r}/\sqrt{L^3}$ basis. 
However, in order to have the $\hat\Gamma_{a,b}$  
operators also diagonal we have to switch to the basis 
\bea
\phi_{a\vec ks}&=&\phi_{b\vec ks}=
\sqrt{\frac{2}{L^3}}\sin\left(\vec k\cdot \vec r\right), \cr
\phi_{a\vec kc}&=&
\phi_{b\vec kc}=\sqrt{\frac{2}{L^3}}\cos\left(\vec k\cdot \vec r\right).
\label{scmodes}
\eea 
Then the Bogoliubov vacuum state in the particle representation reads
\bea
|0_B\ra &\sim& 
\left[\hat a_0^\dagger\hat a_0^\dagger+\sum_{\vec k} 
\lambda_{k}^a\left(\hat a_{\vec ks}^\dagger\hat a_{\vec ks}^\dagger+
\hat a_{\vec kc}^\dagger\hat a_{\vec kc}^\dagger\right)\right]^{N_a/2} \cr
&&\times 
\left[\hat b_0^\dagger\hat b_0^\dagger+\sum_{\vec k} 
\lambda_{k}^b\left(\hat b_{\vec ks}^\dagger\hat b_{\vec ks}^\dagger+
\hat b_{\vec kc}^\dagger\hat b_{\vec kc}^\dagger\right)\right]^{N_b/2} |0\ra,
\cr
&&
\label{finalhomo}
\eea
where 
\bea
\lambda_{k}^a&=&\frac{u_{k,+}^av_{k,+}^a+u_{k,-}^av_{k,-}^a}
{\left(v_{k,+}^a\right)^2+\left(v_{k,-}^a\right)^2+1}
, \cr
\lambda_{k}^b&=&\frac{u_{k,+}^bv_{k,+}^b+u_{k,-}^bv_{k,-}^b}
{\left(v_{k,+}^b\right)^2+\left(v_{k,-}^b\right)^2+1}
,
\label{lambdas}
\eea
and the operators $\hat a_{\vec ks}^\dagger$, $\hat a_{\vec kc}^\dagger$
$\hat b_{\vec ks}^\dagger$ and $\hat b_{\vec kc}^\dagger$ create atoms in the 
modes (\ref{scmodes}).

In the case of the infinite box (i.e. for $L\rightarrow \infty$) if 
$g^2>g_ag_b$ uniform solutions of the Gross-Pitaevskii equations 
become unstable
--- mixing of the $a$ and $b$ components is not energetically favorable 
and the phase separation occurs \cite{homog,timm98,timm03}. 
It manifests itself in the appearance of an imaginary eigenvalue in the 
Bogoliubov spectrum (\ref{hbogspec}). In the case of a finite box the 
minimal value of the 
momentum becomes $2\pi/L$ and the condition for the phase separation is modified,
\be
	g^2 > \left(\frac{\hbar^2 \pi^2}{m_{a} L^2} \frac{1}{\rho_a} + g_a \right)
		\left( \frac{\hbar^2 \pi^2}{m_{b} L^2} \frac{1}{\rho_b} + g_b \right),
\label{pscondh}		
\ee
which shows that for a finite system the minimal value of the parameter 
$g$ leading to the phase separation has to be greater than the corresponding
value for $L\rightarrow \infty$. 

\begin{figure}
\centering
\includegraphics*[width=8.6cm]{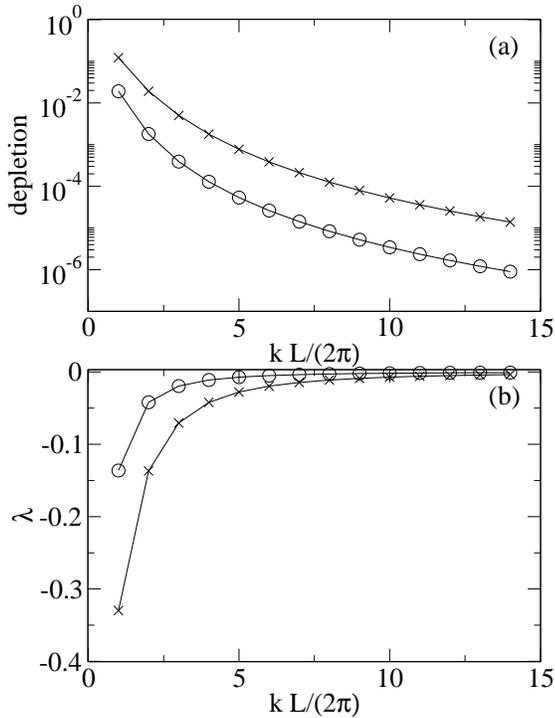}
\caption{
Panel (a) presents condensate depletion, i.e. number of atoms depleted from a 
condensate wavefunction, 
$dN_k^{a,b}=\left(v_{k,+}^{a,b}\right)^2+\left(v_{k,-}^{a,b}\right)^2$, 
to modes (\ref{scmodes}). Panel (b) shows the corresponding values of 
the parameters $\lambda_k^{a,b}$, Eqs.~(\ref{lambdas}). Circles are related to 
the BEC component $a$ while crosses to the component $b$. The results
correspond to the parameters far away from 
the phase separation, i.e. $N_a=5000$, $N_b=20000$, $L=50\;\mu$m,
$a_a=108.8a_0$, $a_b=109.1a_0$ and $a_{ab}=10.0a_0$,  
where $a_0$ is the Bohr radius.
}
\label{hone}
\end{figure}

\begin{figure}
\centering
\includegraphics*[width=8.6cm]{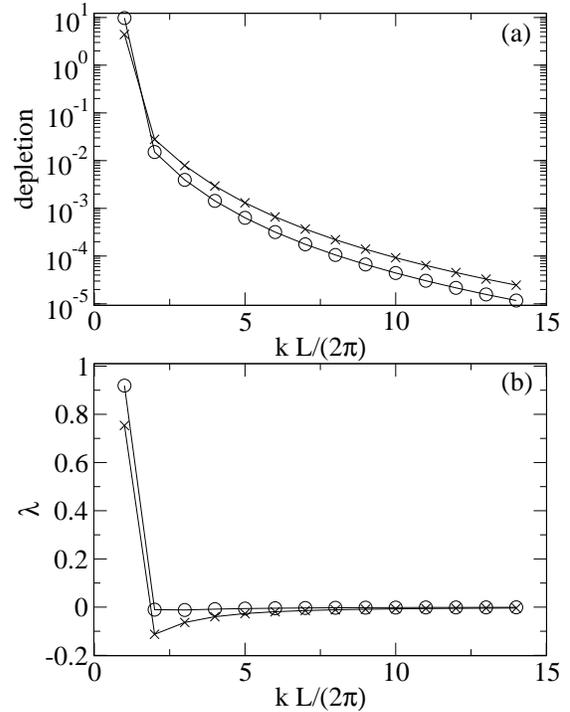}
\caption{ 
The same as Fig.~\ref{hone} but for conditions close to the phase separation,
i.e. $a_{ab}=193.9a_0$. 
}
\label{htwo}
\end{figure}

\begin{figure}
\centering
\includegraphics*[width=8.6cm]{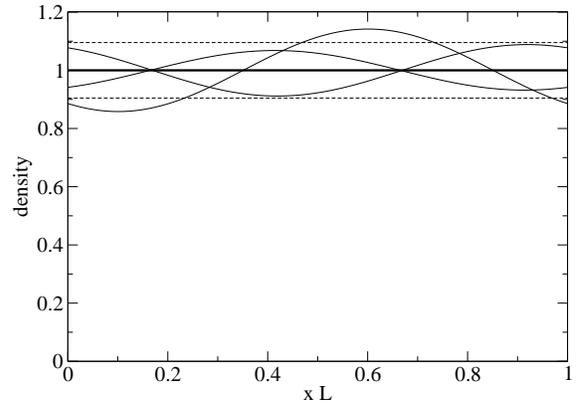}
\caption{  
Solid curves show predictions for results of density measurements of atoms 
belonging to the $a$ component --- the densities integrated over $y$ and $z$ 
directions, i.e. $L\int  \sigma_a(\vec r)\;dydz$, are shown. Thick solid line 
denotes 
the density averaged over many experimental realizations (i.e. the reduced 
single particle density) while dashed lines indicate the average density 
plus/minus standard deviation that is equal to 
$2\sqrt{\Lambda_{k}^a/N_a} \approx 0.1$ (where $k=2\pi/L$).
}
\label{hthree}
\end{figure}

In Fig.~\ref{hone} we show average numbers of atoms depleted to the modes
(\ref{scmodes}) and values of the corresponding $\lambda^{a,b}_k$, 
Eqs.~(\ref{lambdas}), far from the phase separation.
The data correspond to a mixture of $^{87}$Rb atoms in 
two different internal states, $N_a=5000$, $N_b=20000$, $L=50\;\mu$m,
$a_a=108.8a_0$, $a_b=109.1a_0$ and $a_{ab}=10a_0$,
where $a_0$ is the Bohr radius \cite{burke}. The value of the latter scattering length 
can be adjusted by means of a Feshbach resonance \cite{timm98}. 
To obtain predictions for atomic density measurements one has to change phases 
of the modes, see (\ref{phases}), which in the present case of all negative 
$\lambda^{a,b}_k$ leads to
\bea
\varphi_{a\vec ks} &=& \phi_{a\vec ks} e^{-i{\rm Arg}(\lambda^a_\alpha)/2}
=i\phi_{a\vec ks},
\cr
\varphi_{b\vec ks} &=& \phi_{b\vec ks} e^{-i{\rm Arg}(\lambda^b_\alpha)/2}
=i\phi_{b\vec ks},
\eea
and similarly for the $\varphi_{a\vec kc}$ and $\varphi_{b\vec kc}$ modes. 
Due to the fact that $\phi_{a0}$ and $\phi_{b0}$ are real and all the 
$\varphi$ modes are purely
imaginary we obtain, see (\ref{sigmy}),
\bea
\sigma_a(\vec r) &\sim& \phi^2_{a0}(\vec r) + 
\frac{1}{N_a}\left|\sum_{\alpha=1}^{M_a}
q_{a\alpha}\;\varphi_{a\alpha}(\vec r)\right|^2, \cr
\sigma_b(\vec r) &\sim& \phi^2_{b0}(\vec r) + 
\frac{1}{N_b}\left|\sum_{\alpha=1}^{M_b}
q_{b\alpha}\;\varphi_{b\alpha}(\vec r)\right|^2.
\label{przed}
\eea
Because $q_{a\alpha}^2\sim {\Lambda^a_\alpha/2}$,
$q_{b\alpha}^2\sim {\Lambda^b_\alpha}/2$ and $\Lambda^{a,b}_\alpha\ll N_{a,b}$
the density fluctuations turn out to be negligible, i.e. the density remains
almost perfectly flat.

Figure~\ref{htwo} shows similar data as Fig.~\ref{hone} but for $a_{ab}=193.9a_0$,
i.e. chosen 
so that the phase separation condition in an infinite system would be already
fulfilled but it is still not fulfilled in the case of the finite box. 
The numbers of atoms depleted are not dramatically 
greater than the ones in the case
considered previously but now some values of $\lambda^{a,b}_k$ become positive.
The latter has dramatic consequences for density fluctuations 
because the modes $\varphi$ corresponding to the positive 
$\lambda^{a,b}_k$ are real and their contributions to the atomic density 
are of order of $\sqrt{\Lambda^{a,b}_\alpha/N_{a,b}}$.
Indeed, the predictions for atomic density measurements (neglecting 
contributions of order of $\Lambda^{a,b}_\alpha/N_{a,b}$) show that
\bea
\sigma_a(\vec r) &\sim& \phi^2_{a0}(\vec r) + 
\frac{2\phi_{a0}(\vec r)}{\sqrt{N_a}}{\sum_{\vec k}}'\left[
q_{a\vec k s}\varphi_{a\vec k s}(\vec r)+
q_{a\vec k c}\varphi_{a\vec k c}(\vec r)
\right] , \cr
\sigma_b(\vec r) &\sim& \phi^2_{b0}(\vec r) + 
\frac{2\phi_{b0}(\vec r)}{\sqrt{N_b}}{\sum_{\vec k}}'\left[
q_{b\vec k s}\varphi_{b\vec k s}(\vec r)+
q_{b\vec k c}\varphi_{b\vec k c}(\vec r)
\right], \cr &&
\label{po}
\eea
where ${\sum_{\vec k}}'$ runs over modes corresponding to positive 
$\lambda^{a,b}_k$ only. In each experimental realization one has to choose 
$q_{a\vec k s}$, $q_{a\vec k c}$, $q_{b\vec k s}$ and $q_{b\vec k c}$
randomly according to the probability density (\ref{probd}).
In Fig.~\ref{hthree} we show a few examples of the simulations
for atoms belonging to the $a$ component together with 
the averaged result (which corresponds to the reduced single
particle density) 
--- the figure presents the densities integrated over $y$ 
and $z$ directions. Despite the small number of atoms depleted to the 
lowest momentum mode ($\sim 0.3\%$) 
the changes of the density are of order of 10\%.
Standard deviations of the largest scale density fluctuations (i.e. 
corresponding to the quasi-particles with 
the momentum $k=2\pi/L$) behave like 
\be
\sqrt{\frac{\Lambda^{a,b}_k}{N_{a,b}}}\sim 
\frac{1}{\left(a_{ab}^{\rm c}-a_{ab}\right)^{1/4}},
\label{skal2}
\ee
where $a_{ab}^{\rm c}$ is the critical value for the phase separation. 
Exactly at the critical point the Bogoliubov theory breaks down, which is
indicated e.g. by the divergence of the fluctuations in Eq.~(\ref{skal2}).
For the parameters chosen in Fig.~\ref{htwo}-\ref{hthree} we are, however, 
sufficiently far away from the critical point so that the predictions on the 
basis of the Bogoliubov theory are reliable.

From the experimental point of view it is important that
the density fluctuates on a scale of the order of $L$.
One may use low resolution in the density
measurements so that statistical fluctuations will be practically eliminated and 
the only density modulations will correspond to the fluctuations considered 
here.

In the example considered, the range of $a_{ab}$ where one deals with positive 
$\lambda^{a,b}_k$ is about 10$\times$Bohr radius, which should be wide enough
to enable experiments with the density fluctuations (\ref{skal2}).

Note that the structure of the Bogoliubov vacuum state (\ref{finalhomo}) 
shows that 
there are no correlations between atoms belonging to the different components. 
Recently in Ref.~\cite{jaksch}, Bogoliubov vacuum of the form of 
(\ref{final}) has been used in a two
component system 
in the case when the inter-species interaction 
is absent, $g=0$, and the components become fully independent. 
Our analysis indicates that the Bogoliubov
vacuum possesses the same form even in the presence of the inter-species
interaction. Of course, the interaction changes the Bogoliubov modes and 
influences the values of $\lambda_{\alpha}^{a,b}$.

\section{Double well}
\label{doublepart}

In the present section we will consider a simple model where there are analytical 
solutions within the Bogoliubov theory both in the miscible and in the phase 
separated regime.

Let us consider a two component Bose-Einstein condensate in a one-dimensional 
symmetric double well potential under an assumption that the Hilbert space of 
the system is restricted to ground states in each well only (i.e. within the two
mode approximation). For experimental
realizations of the double well problem see \cite{dwell}.
The Hamiltonian of the system, if we choose real functions
as the ground states in the two wells, reads
\bea
	\hat H&=& -\frac{\Omega}{2}\left(\hat A_1^\dagger\hat A_2+
		\hat A_2^\dagger\hat A_1+
		\hat B_1^\dagger\hat B_2+
		\hat B_2^\dagger\hat B_1\right) \cr 
		&& + \frac{U}{2}\left[\left(\hat A_1^\dagger\hat A_1\right)^2+
		\left(\hat A_2^\dagger\hat A_2\right)^2+\left(\hat B_1^\dagger\hat B_1\right)^2+
		\left(\hat B_2^\dagger\hat B_2\right)^2\right] \cr
		&&
		+U_{ab}\left(\hat A_1^\dagger\hat A_1\hat B_1^\dagger\hat B_1+
		\hat A_2^\dagger\hat A_2\hat B_2^\dagger\hat B_2\right),
\label{2wham}
\eea
where the $\hat A_1$ ($\hat B_1$) operator annihilates an atom belonging to
the component 
$a$ ($b$) in the first well and the $\hat A_2$ ($\hat B_2$) operator annihilates 
an atom of the $a$ ($b$) component in the other well.
For the calculations we have chosen potential wells situated at $x=-2,2$ and with
such widths that the ground states of the wells are
\bea
	\psi_1(x)&=&\left(\frac{2}{\pi}\right)^{1/4}e^{-(x+2)^2}, \cr
	\psi_2(x)&=&\left(\frac{2}{\pi}\right)^{1/4}e^{-(x-2)^2}.
\eea

The parameter $\Omega$ stands for the frequency of the
tunneling of atoms between the two wells and $U$ and $U_{ab}$ describe
intra- and inter-condensate interactions.
We will consider the case of symmetric BEC components, i.e.
$N\equiv N_a=N_b=1000$,
$\Omega=5000$, $U=1$ and $U_{ab}>0$ (i.e. all interactions are of
repulsive nature), but similar analysis can be easily performed for
$N_a\ne N_b$ and for tunneling frequencies and intra-species
interactions different for both components. 

We would like to mention that in the limit of large attractive
interactions a two-component entanglement
has been found in the system \cite{twocom2w}.

\subsection{Mean field solutions}

\begin{figure}
\centering
\includegraphics*[width=8.6cm]{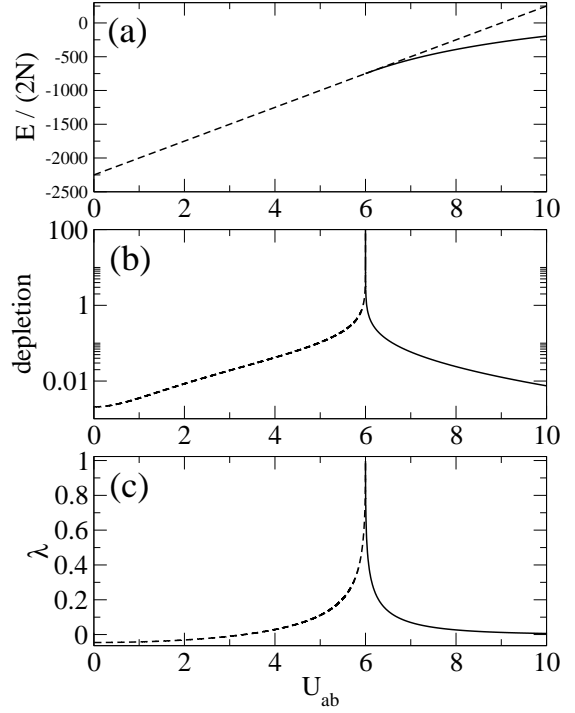}
\caption{  
Panel (a): dashed line shows mean field energy of the symmetric 
Gross-Pitaevskii solution (\ref{cw2w}) while solid line denotes the energy
of the asymmetric solution (\ref{asym}) which appears in the phase separation 
region.
Panel (b) shows depletion of the condensates while panel (c) the corresponding 
values of $\lambda^{a,b}$. Note that, since we consider symmetric interactions,
the depletions are equal for both components and $\lambda^{a}=\lambda^{b}$.
}
\label{one2w}
\end{figure}

For $U_{ab}$ smaller than the critical value 
\be
U^c_{ab}=\frac{\Omega}{N} + U,
\label{uabc}
\ee
the ground state solution of the
Gross-Pitaevskii equations (\ref{GPER}) reveals both condensates symmetrically located in the
double well potential:
\be
	\phi_{a0}(x)=\phi_{b0}(x)=\frac{1}{\sqrt{2}}\left[\psi_1(x)+\psi_2(x)\right],
	\label{cw2w}
\ee
i.e. we are in the miscible regime. However, if the parameters of the system 
fulfil $U_{ab} > U^c_{ab}$
the solution (\ref{cw2w}) is unstable -- the spatial overlap of the
atomic clouds of the different components becomes energetically not favorable 
(see Fig.~\ref{one2w}a) and the phase separation begins. The condensate
wavefunctions are then:
\bea
	\phi_{a0}(x)&=&\alpha\psi_1(x)+\beta\psi_2(x),\cr
	\phi_{b0}(x)&=&\beta\psi_1(x)+\alpha\psi_2(x),
	\label{asym}
\eea
where
\bea
	\alpha&=&\sqrt{\frac{1+\sqrt{1-\gamma^2}}{2}}, \cr
	\beta&=&\sqrt{\frac{1-\sqrt{1-\gamma^2}}{2}},
\eea
and
\be
	\gamma\equiv\frac{\Omega}{N(U_{ab}-U)}.
\ee
Note that in the phase separation regime there are two ground state solutions, 
for exchanging 
$\alpha \leftrightarrow \beta$ in (\ref{asym}) one obtains another 
solution of the Gross-Pitaevskii equations. On the basis of these two solutions
two different Bogoliubov vacuum states can be obtained.
In the following we will
show that sufficiently far away from the critical point the ground 
state of the system can be approximated by preparing a superposition of the two 
Bogoliubov vacuum states.


\subsection{The Bogoliubov vacuum -- miscible regime}

The solution of the Bogoliubov-de~Gennes equations reveals two quasi-particles
corresponding to energies:
\bea
	E_{\pm} &=& \left[ \Omega\left(\Omega + NU \pm N U_{ab}\right) 
	\right]^{1/2},
	\label{qpensym}
\eea
and modes:
\be
\left(
\begin{array}{c}
	u^a_{\pm}(x) \\ 
	v^a_{\pm}(x) \\
	u^b_{\pm}(x) \\
	v^b_{\pm}(x)
\end{array}
\right)= 
\left(
\begin{array}{c}
	(\Omega + 2E_\pm) \phi_{a1}(x) \\ 
	(2E_\pm - \Omega) \phi_{a1}(x) \\
	(\Omega + 2E_\pm) \phi_{b1}(x) \\ 
	(2E_\pm - \Omega) \phi_{b1}(x)
\end{array}
	\right) \chi_\pm,
\ee
where 
\be
	\phi_{a1}(x)=\phi_{b1}(x)=\frac{1}{\sqrt{2}}\left[\psi_1(x)-\psi_2(x)\right]
\ee
and
\bea
\chi_\pm = \frac{1}{4 \sqrt{\Omega E_\pm }},
\eea
is the normalization factor.

The Bogoliubov vacuum state can be written in the form (\ref{final}) 
using
\bea
	\lambda^a = \lambda^b = 
	\frac{\left(4 E^2_+ - \Omega^2\right) \chi^2_+ + \left(4 E^2_- - \Omega^2\right)
	\chi^2_-}
	{1 + (2E_+ - \Omega)^2 \chi^2_+ + (2E_- - \Omega)^2 \chi^2_-}.
\label{lambda2wm}	
\eea
The parameters (\ref{lambda2wm}) and the depletion are shown versus $U_{ab}$
in Fig.~\ref{one2w}.

As can be seen from the quasiparticle excitation energy (\ref{qpensym}), when
the condition $U_{ab} > U^c_{ab}$ is fulfilled, 
the spectrum is no longer real and the solution (\ref{cw2w}) 
becomes unstable. The ground state of the Gross-Pitaevskii equations 
shows separation of the atomic clouds of the $a$ and $b$ components.

\subsection{The Bogoliubov vacuum -- phase separation regime}

Now the quasi-particle excitation energies are
\bea
	E_\pm = \left\{\frac{\Omega}{2} \left[ \frac{\Omega}{2}
		\left( \frac{\alpha}{\beta} + 
		\frac{\beta}{\alpha} \right)^2 + 4 U \alpha\beta \right]
		\pm 2 \Omega U_{ab} \alpha\beta \right\}^{1/2}, \cr
\eea
and the quasiparticle modes, corresponding to the condensate 
wavefunctions (\ref{asym}), are proportional to 
\bea
	\phi_{a1}(x)&=&\beta\psi_1(x)-\alpha\psi_2(x),\cr
	\phi_{b1}(x)&=&\alpha\psi_1(x)-\beta\psi_2(x).
\eea
We skip here rather long expressions for the quasiparticles since they can be easily
obtained with the help of the Bogoliubov transformation. Behaviour of the component 
$a$ (or $b$) depletion and of the parameters $\lambda^a, \lambda^b$ versus 
$U_{ab}$ is depicted in Fig.~\ref{one2w}.

A Hamiltonian of a BEC system in a symmetric double well potential is invariant
under the parity inversion, i.e. if we reverse coordinates of all particles
the Hamiltonian does not change. It implies that eigenstates of our system 
(unless there is a degeneracy) must be also eigenstates of the parity operator. 
In the phase separation regime the Gross-Pitaevskii solutions (\ref{asym}) 
are neither even nor odd functions and the corresponding Bogoliubov vacuum,
\bea
|0_B\ra &\sim& 
\left[\left(\hat a_0^\dagger\right)^2+ 
\lambda^a\left(\hat a_1^\dagger\right)^2\right]^{N/2} \cr
&&\times 
\left[\left(\hat b_0^\dagger\right)^2+
\lambda^b\left(\hat b_1^\dagger\right)^2\right]^{N/2} |0\ra,
\eea
is not an eigenstate of the parity operator. 
Exchanging $\alpha$ with $\beta$ one obtains another Gross-Pitaevskii solution
and another Bogoliubov vacuum state. A proper parity state can be obtained by 
preparing a superposition of the two states
\bea
|0_B\ra + |0_B ({\alpha\leftrightarrow \beta})\ra.
\label{parity}
\eea
The state (\ref{parity}) is a good approximation for the ground state of the system 
if we are not very close to the critical point. That is, if we increase $U_{ab}$ for fixed $N$,
the states $|0_B\ra$ and $|0_B ({\alpha\leftrightarrow \beta})\ra$ 
very quickly become practically orthogonal, and the sooner it takes place, the greater $N$ we choose.
Note that even in the regime of these states being orthogonal, the corresponding mean field states
(\ref{asym}) need not be orthogonal at all.
If we are, however, far away from the critical point also the Gross-Pitaevskii solutions 
have zero overlap,
i.e. $\la \phi_{a0}|\phi_{b0}\ra \approx 0$. Then the state (\ref{parity}) is a
a Schr\"odinger cat state \cite{schcat} 
which is strongly vulnerable to atomic losses --- loss of a small number of atoms 
is sufficient to distort completely the coherent superposition in 
(\ref{parity}).

At the critical point the Bogoliubov theory breaks down because higher order 
terms become dominant.

\subsection{Density fluctuations}

We see in Fig.~\ref{one2w} that approaching the critical point
in the miscible regime the depletion of the condensates and the Bogoliubov 
vacuum parameters $\lambda^{a(b)}$ grow.
Very close to the critical point the depletion is very large and 
the Bogoliubov theory can not be longer applied. However, in the vicinity 
of the point there is a regime where the depletions are very small compared 
to the total particle numbers and the parameters $\lambda^{a(b)}$ are positive.
The modes $\phi_{a0}$, $\phi_{b0}$, $\phi_{a1}$ and $\phi_{b1}$ are real
and the appearance of the positive $\lambda^{a,b}$ indicates (similarly as 
in the previous section) that density fluctuations become considerable, 
i.e. of order of $\sqrt{\Lambda^{a,b}/N}$,
\bea
\sigma_a(x)&\sim&\phi^2_{a0}(x)+2\frac{q_a}{\sqrt{N}}
\phi_{a0}(x)\phi_{a1}(x), \cr
\sigma_b(x)&\sim&\phi^2_{b0}(x)+2\frac{q_b}{\sqrt{N}}
\phi_{b0}(x)\phi_{b1}(x), \cr &&
\label{d2w}
\eea
where $q_a$ and $q_b$ have to be chosen randomly according to 
(\ref{probd}) in order to get predictions for the results of the density
measurement in different experimental realizations.

\begin{figure}
\centering
\includegraphics*[width=8.6cm]{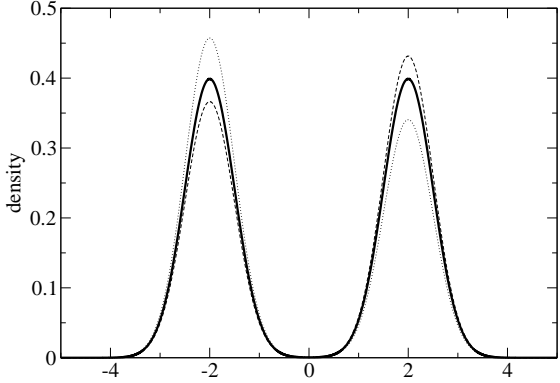}
\caption{
Density fluctuations of the $a$ component in the miscible regime close to the critical point,
i.e. for $U_{ab}=5.995$. 
Solid line corresponds to the mean field solution, dashed and dotted lines are two 
examples of possible realizations of the experiment. The standard deviation of the density
fluctuation is $\sqrt{2\Lambda^a /N} \phi_{a0} \phi_{a1} \approx 0.09 \phi_{a0} \phi_{a1} $.
}
\label{two2w}
\end{figure}

A few examples of the density measurements of the $a$ component atoms 
in the miscible regime are shown in Fig.~\ref{two2w} for 
$U_{ab}^{\rm c}-U_{ab}=0.005$. 
Standard deviations of the density fluctuations behave like:
\be
	\sqrt{\frac{\Lambda^{a}}{N}} \sim 
	\frac{1}{\left(U_{ab}^{\rm c}-U_{ab}\right)^{1/4}},
	\label{skal1dw}
\ee
i.e. similarly as in the homogeneous case, see (\ref{skal2}).

On the other side of the critical point, i.e. in the phase separation regime, 
we deal with a state of the form (\ref{parity}) which
for $U_{ab}^{\rm c}-U_{ab}=-0.005$ and $N=5000$ is a good approximation 
for the ground state of the system, indeed
$|\la 0_B|0_B ({\alpha\leftrightarrow \beta})\ra|^2$ is of order of $10^{-8}$.
One may also expect (similarly as in the miscible regime)
substantial density fluctuations. However, for a superposition of the Bogoliubov vacuum states
(\ref{parity}) we cannot simulate density measurements with the help of the method
described in Sec.~III.

\section{Conclusions}

We have considered a number conserving version of the Bogoliubov theory for a
two component Bose-Einstein condensate, with the fixed number of atoms in each 
component. We have shown that the Bogoliubov vacuum state can 
be written in the particle representation in a simple form, provided that eigenstates
of the reduced single particle density matrices diagonalize the operators 
(\ref{gammaoper}). Having the 
Bogoliubov vacuum in the particle representation one can easily obtain
predictions for density measurements in single experiments. 

The introduced formalism has been applied to the analysis of a two component 
homogeneous condensate and a two component condensate in a double well potential.
In finite homogeneous systems, when parameters of the system approach a phase 
separation condition, considerable density fluctuations appear before the system 
becomes unstable. This behaviour is different than in infinite systems,
where the phase separation happens abruptly.
The range of the parameter values where the substantial 
fluctuations are observed indicates that the results presented here can be verified 
experimentally.

In the case of condensates in a double well potential we are able to
describe the system in a vicinity of the critical point 
both in the miscible condensates regime and in the phase 
separation region. Considerable density fluctuations can be expected 
if the parameters approach the critical values.

\section*{ Acknowledgements }  
We are grateful to Roman Marcinek and 
Kuba Zakrzewski for critical reading of the manuscript.
The work of BO was supported by Polish Government scientific funds 
(2005-2008) as a research project. KS was supported by the KBN grant 
PBZ-MIN-008/P03/2003. 
Supported by Marie Curie ToK project COCOS (MTKD-CT-2004-517186).

\appendix
\setcounter{section}{0}\setcounter{subsection}{0}
\setcounter{equation}{0}
\renewcommand{\theequation}{A-\arabic{equation}}
\section*{Appendix A}
\label{appendixA}

We begin with a short reminder of the results of the number conserving 
version of the Bogoliubov theory. Following \cite{sorensen}
we will perform the perturbation expansion of the hamiltonian.
The decomposition (\ref{decompR}) allows us to expand the
Hamiltonian in powers of {\it small} operators  $\delta\hat\psi_a$ 
and $\delta\hat\psi_b$. As mentioned in Sec.~\ref{BVS}, 
minimizing the energy of the system in the zero order we obtain coupled Gross-Pitaevskii 
equations (\ref{GPER}) that allow us to find the condensate wavefunctions
$\phi_{a0}$ and $\phi_{b0}$. 
The first order terms of the Hamiltonian disappear and in the second order we 
obtain an effective Hamiltonian
\be
\hat H_{\rm eff}\approx \frac12 \int d^3r
\left(\hat\Lambda_a^\dagger,-\hat\Lambda_a,\hat\Lambda_b^\dagger,
-\hat\Lambda_b\right) {\cal L}
\left(\begin{array}{c} 
\hat\Lambda_a \\ 
\hat\Lambda_a^\dagger \\
\hat\Lambda_b \\
\hat\Lambda_b^\dagger
\end{array}\right),
\label{effapp}
\ee 
where
\begin{widetext}
\be
{\cal L}=\left(
\begin{array}{cccc}
	H_{GP}^a+g_a N_a \hat Q_a|\phi_{a0}|^2\hat Q_a &  g_a N_a\hat Q_a\phi_{a0}^{2}\hat Q_a^* &  
		g \sqrt{N_a N_b} Q_a \phi_{a0} \phi^*_{b0} Q_b &  g \sqrt{N_a N_b} Q_a \phi_{a0}\phi_{b0} Q^*_b \\
	-g_a N_a\hat Q^*_a\phi_{a0}^{*2}\hat Q_a & -H_{GP}^a-g_a N_a \hat Q^{*}_a|\phi_{a0}|^2\hat Q^{*}_a &
		-g \sqrt{N_a N_b} Q^*_a \phi^{*}_{a0}\phi^*_{b0} Q_b & -g \sqrt{N_a N_b} Q^*_a \phi^*_{a0}\phi_{b0} Q^*_b \\
	 g\sqrt{N_a N_b} Q_b \phi^*_{a0}\phi_{b0} Q_a &  g\sqrt{N_a N_b} Q_b \phi_{a0}\phi_{b0} Q^*_a &
	 		H_{GP}^b+g_b N_b \hat Q_b|\phi_{b0}|^2\hat Q_b & g_b N_b\hat Q_b\phi_{b0}^{2}\hat Q_b^* \\
	 -g\sqrt{N_a N_b} Q^*_b \phi^*_{a0}\phi^*_{b0} Q_a & -g\sqrt{N_a N_b} Q^*_b \phi_{a0}\phi^*_{b0} Q^*_a  &
	 		-g_b N_b\hat Q^*_b\phi_{b0}^{*2}\hat Q_b & -H_{GP}^b-g_b N_b \hat Q^{*}_b|\phi_{b0}|^2\hat Q^{*}_b\\
\end{array}
\right),
\label{Lop}
\ee
\end{widetext}
and
\bea
\hat Q_a&=&1-|\phi_{a0}\ra\la\phi_{a0}|, \cr
\hat Q_b&=&1-|\phi_{b0}\ra\la\phi_{b0}|.
\eea
The $\hat\Lambda_a(\vec r)$ and $\hat\Lambda_b(\vec r)$ operators
(\ref{Lambdaop}) fulfil the following commutation relations
\bea
[\hat\Lambda_a(\vec r),\hat\Lambda_a^\dagger(\vec r\;')]\approx 
\la \vec r|\hat Q_a|\vec r\;'\ra, \cr
[\hat\Lambda_b(\vec r),\hat\Lambda_b^\dagger(\vec r\;')]\approx 
\la \vec r|\hat Q_b|\vec r\;'\ra.
\eea
Note that action of the $\hat\Lambda_a$ and $\hat\Lambda_b$ operators preserves
numbers of atoms in the system.
Diagonalization of the effective Hamiltonian amounts to solving the 
eigen-equation for the non-hermitian operator $\cal L$ (i.e. the
Bogoliubov-de~Gennes equations).

The $\cal L$ operator possesses two symmetries (similarly to the 
symmetries of the original Bogoliubov-de~Gennes equations \cite{castin}),
\bea
u_1{\cal L}u_1&=&-{\cal L}^*, \cr
u_3{\cal L}u_3&=&{\cal L}^\dagger,
\label{symmL}
\eea
where
\bea
u_1=\left(
\begin{array}{c|c}
 \sigma_1 & 0 \\
\hline
 0 & \sigma_1 \\
\end{array}
\right),
&&
u_3=\left(
\begin{array}{c|c}
 \sigma_3 & 0 \\
\hline
 0 & \sigma_3 \\
\end{array}
\right),
\eea
and 
\bea
\sigma_1=\left(\begin{array}{cc}
0&1\\
1&0
\end{array}\right), &&
\sigma_3=\left(\begin{array}{cc}
1&0\\
0&-1
\end{array}\right),
\eea
are the first and third Pauli matrices, respectively. 
Suppose that all eigenvalues of the ${\cal L}$ operator are real. 
The symmetries (\ref{symmL}) imply that if
\be
|\Psi_n^{\rm R}\rangle=
\left(
\begin{array}{c}
|u_n^a\rangle  \\
|v_n^a\rangle \\
|u_n^b\rangle  \\
|v_n^b\rangle \\
\end{array}
\right),
\ee
is a right eigenvector of the ${\cal L}$ with eigenvalue $E_n$, then 
$|\Psi_n^{\rm L}\rangle=u_3|\Psi_n^{\rm R}\rangle$
is a left
eigenvector of the same eigenvalue $E_n$, and 
$u_{1} |\Psi^{R\ast}_{n}\rangle$
is a
right eigenvector with eigenvalue $-E_n$.  

There are four eigenvectors of 
${\cal L}$ corresponding to a zero eigenvalue,
\bea
\left(
\begin{array}{c}
|\phi_a \ra \\
0 \\
0  \\
0 \\
\end{array}
\right),\;
\left(
\begin{array}{c}
0  \\
|\phi_a^*\ra \\
0  \\
0 \\
\end{array}
\right),\;
\left(
\begin{array}{c}
0  \\
0 \\
|\phi_b \ra \\
0 \\
\end{array}
\right),\;
\left(
\begin{array}{c}
0 \\
0 \\
0  \\
|\phi_b^*\ra \\
\end{array}
\right).
\eea
The other eigenstates of the ${\cal L}$ operator we divide into two 
families "+" and "$-$", according to
\be
\langle\Psi_n^{\rm R}| u_3|\Psi_{n'}^{\rm R}\rangle=\pm \delta_{n,n'}.
\ee
Having the complete set of the eigenvectors of the 
$\cal L$ we obtain an important
completeness relation
\begin{widetext}
\bea
\hat 1
&=& \sum_{n\in"+"}
\left(
\begin{array}{c}
|u_n^a \ra \\
|v_n^a\ra \\
|u_n^b\ra \\
|v_n^b\ra
\end{array}
\right) 
\left(\la u_n^a |, -\la v_n^a|, \la u_n^b|, -\la v_n^b| \right)
+ \sum_{n\in"+"}
\left(
\begin{array}{c}
|v_n^{a*} \ra \\
|u_n^{a*}\ra \\
|v_n^{b*}\ra \\
|u_n^{b*}\ra
\end{array}
\right) 
\left(-\la v_n^{a*} |, \la u_n^{a*}|, -\la v_n^{b*}|, \la u_n^{b*}| \right)
\cr
&&+
\left(
\begin{array}{cccc}
|\phi_{a0}\ra\la \phi_{a0}| & 0 & 0 & 0 \\
0 & |\phi_{a0}^*\ra\la \phi_{a0}^*| & 0 & 0 \\
0 & 0 & |\phi_{b0}\ra\la \phi_{b0}| & 0 \\
0 & 0 & 0 & |\phi_{b0}^*\ra\la \phi_{b0}^*| \\
\end{array}
\right). 
\label{compr}
\eea
\end{widetext}
The eigenvectors of the $\cal L$ operator define the Bogoliubov transformation
\be
\left(\begin{array}{c} 
\hat\Lambda_a \\ 
\hat\Lambda_a^\dagger \\
\hat\Lambda_b \\
\hat\Lambda_b^\dagger
\end{array}\right)=
\sum_{n\in"+"}
\left(
\begin{array}{c}
u_n^a  \\
v_n^a \\
u_n^b \\
v_n^b
\end{array}
\right)\hat c_n +
\sum_{n\in"+"}
\left(
\begin{array}{c}
v_n^{a*} \\
u_n^{a*} \\
v_n^{b*} \\
u_n^{b*}
\end{array}
\right)
\hat c_n^\dagger ,
\label{BTexp}
\ee
where the quasi-particle operators (\ref{boper})
fulfill the bosonic commutation relation
$[ \hat{c}_{n}, \hat{c}^{\dagger}_{n'} ] \approx \delta_{n,n'}$. 
Employing the Bogoliubov transformation we obtain the effective 
Hamiltonian in a diagonal form (\ref{efffin}).

\appendix
\setcounter{section}{0}\setcounter{subsection}{0}
\setcounter{equation}{0}
\renewcommand{\theequation}{B-\arabic{equation}}
\section*{Appendix B}
\label{appendixB}

The Bogoliubov vacuum state $|0_B\ra$ is an eigenstate of the effective
Hamiltonian (\ref{efffin}) that is annihilated by all quasi-particle annihilation operators.
Let us show that the Bogoliubov vacuum can be obtained from the particle 
vacuum by applying some particle creation operators 
$\hat d_a^\dagger$ and $\hat d_b^\dagger$
\be
|0_B\ra \sim \left(\hat d_a^\dagger\right)^{M_a}
\left(\hat d_b^\dagger\right)^{M_b}|0\ra,
\label{ddBog}
\ee
where we require that 
$\hat d_a^\dagger$ and $\hat d_b^\dagger$ commute with all quasi-particle 
annihilation operators \cite{dziarmaga},
\bea
[\hat c_n,\hat d_a^\dagger]&=&0, \cr
[\hat c_n,\hat d_b^\dagger]&=&0.
\label{commute}
\eea
Then the state (\ref{ddBog}) is indeed annihilated by all quasi-particle 
annihilation operators,
\be
\hat c_n\left(\hat d_a^\dagger\right)^{M_a}
\left(\hat d_b^\dagger\right)^{M_b}|0\ra
=\left(\hat d_a^\dagger\right)^{M_a}
\left(\hat d_b^\dagger\right)^{M_b}\hat c_n|0\ra=0.
\ee

The set of equations (\ref{commute}) is solved by
the particle creation operators in the form
\cite{dziarmaga,ansatzL}
\bea
\hat d_a^\dagger&=&\hat a_0^\dagger\hat a_0^\dagger+
\sum_{\alpha,\beta=1}^\infty Z_{\alpha\beta}^a\hat a_\alpha^\dagger\hat a_\beta^\dagger \cr
\hat d_b^\dagger&=&\hat b_0^\dagger\hat b_0^\dagger+
\sum_{\alpha,\beta=1}^\infty Z_{\alpha\beta}^b\hat b_\alpha^\dagger\hat b_\beta^\dagger,
\label{ans}
\eea
where $\hat a_\alpha^\dagger$ ($\hat b_\alpha^\dagger$) are bosonic particle creation 
operators that create atoms in modes $\phi_{a\alpha}$ ($\phi_{b\alpha}$) orthogonal
to the condensate wavefunction $\phi_{a0}$ ($\phi_{b0}$). $Z_{\alpha\beta}^a$ 
and $Z_{\alpha\beta}^b$ are symmetric matrices to be found.

Substituting the ansatz (\ref{ans}) into (\ref{commute}) we obtain equations:
\bea
\la v_n^a|\phi_{a\alpha}^*\ra&=&\sum_{\beta=1}^\infty 
\la u_n^a |\phi_{a\beta}\ra Z_{\beta\alpha}^a \cr
\la v_n^b|\phi_{b\alpha}^*\ra&=&\sum_{\beta=1}^\infty 
\la u_n^b |\phi_{b\beta}\ra Z_{\beta\alpha}^b, 
\eea
which, when multiplied by $\la \phi_{a\gamma}|u_n^a \ra$ and 
$\la \phi_{b\gamma}|u_n^b \ra$, respectively, and summed over $n$, 
are transformed to
\bea
\la \phi_{a\gamma}|\hat\Gamma_a|\phi_{a\alpha}^*\ra &=&\sum_{\beta=1}^\infty 
\la \phi_{a\gamma}| \hat U_a |\phi_{a\beta}\ra Z_{\beta\alpha}^a \cr
\la \phi_{b\gamma}|\hat\Gamma_b|\phi_{b\alpha}^*\ra &=&\sum_{\beta=1}^\infty 
\la \phi_{b\gamma}| \hat U_b |\phi_{b\beta}\ra Z_{\beta\alpha}^b,
\eea
where
\bea
\hat U_a=\sum_{n\in"+"}|u_n^a\ra\la u_n^a|, \quad
\hat U_b=\sum_{n\in"+"}|u_n^b\ra\la u_n^b|,
\eea
and $\hat\Gamma_{a,b}$ are defined in (\ref{gammaoper}).
The completeness relation (\ref{compr}) implies that the $\hat\Gamma_a$
and $\hat\Gamma_b$ operators are symmetric and that
\bea
\hat U_a&=&\sum_{n\in"+"}|v_n^{a*}\ra\la v_n^{a*}|+\hat 1_\perp^a, 
\label{ua1} \\
\hat U_b&=&\sum_{n\in"+"}|v_n^{b*}\ra\la v_n^{b*}|+\hat 1_\perp^b,
\eea
where $\hat 1_\perp^a$ and $\hat 1_\perp^b$ are the identity operators in the
subspaces orthogonal to the condensate wavefunctions $\phi_{a0}$ and
$\phi_{b0}$, respectively. Comparing Eq.~(\ref{ua1}) with 
\bea
\la 0_B|\hat\psi_a^\dagger(\vec r) \hat\psi_a(\vec r\;')|0_B\ra
&=& N_a\phi_{a0}^*(\vec r)\phi_{a0}(\vec r\;') \cr
&&+\sum_{n\in"+"}
v_n^a(\vec r)v_n^{a*}(\vec r\;').
\eea 
we see that the $\hat U_{a}$ operator is a sum of a part of the reduced 
single particle density operator corresponding to the subspace orthogonal to the
condensate wavefunction $\phi_{a0}$ and the identity operator 
$\hat 1_\perp^a$. Similar statement is true in the
case of the $b$ component. Thus, if we choose as a basis $\phi_{a\alpha}$
($\phi_{b\alpha}$), the eigenstates of the single particle density matrix,
we get the $\hat U_{a}$ ($\hat U_{b}$) operator in a diagonal form. Then one 
obtains immediately the solutions for the $Z^{a,b}_{\alpha\beta}$ matrices, i.e.
\bea
Z_{\alpha\beta}^a=
\frac{\la \phi_{a\alpha}|\hat\Gamma_a|\phi_{a\beta}^*\ra}{dN^a_\alpha+1}, 
\quad
Z_{\alpha\beta}^b=
\frac{\la \phi_{b\alpha}|\hat\Gamma_b|\phi_{b\beta}^*\ra}{dN^b_\alpha+1},
\eea
where $dN_\alpha^{a,b}$ are the eigenvalues of the single particle density
matrices, that is numbers of atoms depleted from the condensate wavefunctions 
to other eigenmodes.
The $Z^{a}_{\alpha\beta}$,  $Z^{b}_{\alpha\beta}$,
$\la \phi_{a\alpha}|\hat\Gamma_{a}|\phi_{a\beta}^*\ra$ and
$\la \phi_{b\alpha}|\hat\Gamma_{b}|\phi_{b\beta}^*\ra$ 
matrices are symmetric. Thus, in order the ansatz (\ref{ans}) to be
self-consistent the $\hat\Gamma_{a,b}$ 
operators have to be also diagonal in the basis of 
the eigenvectors of the single particle density matrices.
In Sec.~\ref{homopart} and \ref{doublepart} we show examples 
where indeed this is the case.
Final form of the solution for the Bogoliubov vacuum state in the
particle representation is presented in (\ref{final}).


\end{document}